\documentstyle[psfig,conf_iap,]{article}

\newcommand{\DMR}{DMR}

\newcommand{\Boomerang}{BOOMERANG}
\newcommand{\Maxima}{MAXIMA}
\newcommand{\DASI}{DASI}

\newcommand{\cbi}{CBI}

\def\PRD{{Phys. Rev.} D}

\def\apj{{Ap.J.}}

\begin{document}
\heading{%
%
CMB observations with the Cosmic Background Imager (CBI) Interferometer
%
} 
\par\medskip\noindent
\author{%
C.~R.~Contaldi${}^1$, J.~R.~Bond${}^1$, D.~Pogosyan${}^2$, B.~S.~Mason${}^{3,4}$,
S.~T.~Myers${}^5$, T.~J.~Pearson${}^3$, U.~L.~Pen${}^1$, S.~Prunet${}^{1,6}$, A.~C.~Readhead${}^3$,
M.~I.~Ruetalo${}^{1,7}$, J.~L.~Sievers${}^3$, J.~W.~Wadsley${}^8$, P.~J.~Zhang${}^{1,7}$
}
\address{
Canadian Institute for Theoretical Astrophysics 60 St. George
Street,  Toronto Ontario M5S 3H8\\ 
${}^2$ University of Alberta, ${}^3$ Caltech, ${}^4$ NRAO, ${}^5$ IAP, ${}^6$ University of Toronto, ${}^7$ McMaster University
}

\begin{abstract}
We review the recently published results from the CBI's first season
of observations. Angular power spectra of the CMB were obtained from
deep integrations of 3 single fields covering a total of 3 deg$^2$
and 3 shallower surveys of overlapping (mosaiced) fields covering a
total of 40 deg$^2$. The observations show a damping of the
anisotropies at high-$\ell$ as expected from the standard scenarios of
recombination. We present parameter estimates obtained from the data
and discuss the significance of an excess at $\ell>2000$ observed in
the deep fields.
\end{abstract}

\section{Introduction}
 The \cbi\ is an array of 13 0.9-meter diameter Cassegrain antennas
mounted on a 6-meter diameter tracking platform. The results presented
here are based on an analysis of data obtained from January through to
December 2000 \cite{Mason02b,Pearson02}. Previous experiments such as
\Boomerang\cite{Netterfield01}, \DASI\cite{DASI01} and
\Maxima\cite{Maxima01} have measured precisely the shape of a first
peak at $\ell \sim 220$, consistent with an $\Omega_{tot}=1$
$\Lambda$CDM universe with adiabatic, inflationary seeded
perturbations. A significant detection of a second peak has also been
established \cite{deBernardis01} with evidence for a third. The \cbi\
observations have now extended the multipole range observed by multiple
band experiments by a factor of $3$. The results have confirmed
another important element of the adiabatic, inflationary paradigm, the
damping at high multipoles due to the viscous drag over the finite
width of the last scattering surface \cite{Sievers02}. Mason et
al.(2002) \cite{Mason02b} give a description of the \cbi\ instrumental
setup and observing strategy. Here we review the main results and
cosmological parameter fits obtained from the observations and discuss
the nature of the possible excess observed on the smallest angular
scales at $2000<\ell<4000$.

\section{Results}
The instrument observes in 10 frequency channels
spanning the band $26-36$GHz and measures 78 baselines
simultaneously. Our power spectrum estimation pipeline is described in
\cite{Myers02} and involves an optimal compression of the ${\cal
O}(10^5)$ visibility measurements of each field into a coarse grained
lattice of visibility estimators. Known point sources are projected
out of the data sets when estimating the primary anisotropy spectrum by
using a number of constraint matrices. The positions are obtained from
the (1.4GHz) NVSS catalog \cite{Condon98}. When projecting out the
sources we use large amplitudes which effectively marginalize over all
affected modes. This insures robustness with respect to errors in the
assumed fluxes of the sources. The residual contribution of sources
below our $S_{1.4}=3.4$ mJy cutoff is treated as a white noise background
with an estimated amplitude of $0.08\pm 0.04$ Jy/sr${^-1}$
\cite{Mason02b}.

\begin{figure}\label{fig:spectra}
\begin{tabular}{cc}
\vbox{
\psfig{figure=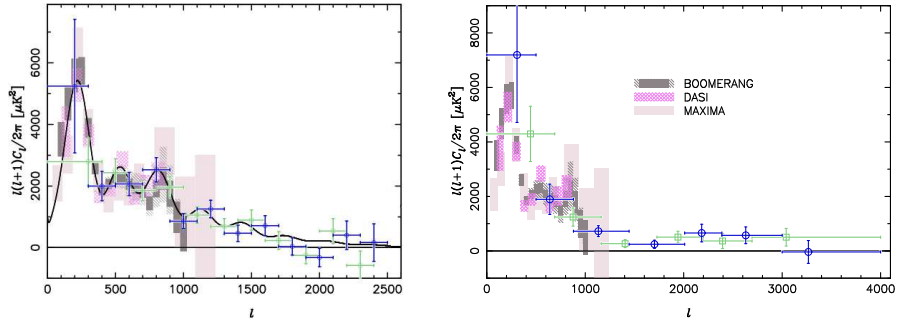,height=4.2cm}
}&
\vbox{
\psfig{figure=cbi_fig1b.ps,height=4.2cm}
}
\end{tabular}

\caption[]{\small The mosaic (left) and deep (right) field power spectra. Two separate binnings of the data are shown (blue and green). We also include 1-sigma confidence intervals for the \Boomerang, \DASI\ and \Maxima\ experiments. The hatched (grey) errors show the effect of the beam uncertainty on the \Boomerang\ measurements. The solid curve in the left panel is a best fit spectrum to all the data shown.
}
\end{figure}

The power spectra for both the mosaic and deep observations are shown in Figure~\ref{fig:spectra}. Both show a clear detection of the expected damping of the power at $\ell>1000$. 

We fit a minimal set of inflationary parameters to the data; $\omega_b
= \Omega_bh^2$, $\omega_{\rm cdm}=\Omega_{\rm cdm}h^2$, $\Omega_{\Lambda}$,
 $\Omega_{tot}$, $n_s$, $\tau_c$. We also include calibration errors and
beam errors where applicable. Our parameter fitting pipeline is
described in detail in Sievers et al. (2002) \cite{Sievers02}. The
results for the combination of \cbi\ and \DMR\ data are shown in
Table~\ref{tab:param} for various combinations of priors. Our
parameter fits and best fit models are consistent with previous
results from fits to data at lower multipoles. We have also carried
out a combined analysis of the \cbi\ data with the \Boomerang,
\Maxima\ and \DASI\ data and with a compilation of data predating
April 2001 \cite{Sievers02}.

\begin{center}
\begin{table}[th]
\caption{Parameter constraints from CBI+DMR for a selection of priors.\label{tab:param}}
\vspace{0.4cm}
\begin{tabular}{|l|cccccc|}
\hline
{Priors}
&{$\Omega_{tot}$}
&{$n_s$}
&{$\Omega_bh^2$}
&{$\Omega_{\rm cdm}h^2$}
&{$\Omega_{\Lambda}$}
&{$\tau_c$}\\
\hline
{\small wk}
& $0.99^{0.12}_{0.12}$ 
& $1.05^{0.09}_{0.08}$ 
& $0.022^{0.015}_{0.009}$ 
& $0.17^{0.08}_{0.06}$ 
& $0.40^{0.25}_{0.27}$ 
& $0.22^{0.19}_{0.16}$ 
\\
{\small wk+LSS}
& $1.01^{0.09}_{0.06}$ 
& $1.02^{0.11}_{0.07}$ 
& $0.026^{0.014}_{0.010}$ 
& $0.12^{0.03}_{0.03}$ 
& $0.64^{0.11}_{0.14}$ 
& $0.14^{0.22}_{0.11}$ 
\\
{\small wk+SN}
& $1.02^{0.09}_{0.08}$ 
& $1.07^{0.09}_{0.09}$ 
& $0.027^{0.015}_{0.011}$ 
& $0.12^{0.05}_{0.05}$ 
& $0.70^{0.08}_{0.09}$ 
& $0.24^{0.18}_{0.18}$ 
\\
{\small wk+LSS+SN}
& $1.00^{0.10}_{0.06}$ 
& $1.06^{0.09}_{0.08}$ 
& $0.027^{0.014}_{0.011}$ 
& $0.12^{0.04}_{0.04}$ 
& $0.70^{0.07}_{0.07}$ 
& $0.21^{0.20}_{0.15}$ 
\\\hline
{\small Flt+wk}
& (1.00) 
& $1.04^{0.10}_{0.08}$ 
& $0.023^{0.010}_{0.008}$ 
& $0.15^{0.06}_{0.04}$ 
& $0.46^{0.22}_{0.29}$ 
& $0.22^{0.19}_{0.16}$ 
\\
{\small Flt+wk+LSS}
& (1.00) 
& $1.01^{0.10}_{0.07}$ 
& $0.025^{0.010}_{0.008}$ 
& $0.13^{0.02}_{0.01}$ 
& $0.64^{0.10}_{0.13}$ 
& $0.15^{0.17}_{0.11}$ 
\\
{\small Flt+wk+SN}
& (1.00) 
& $1.06^{0.11}_{0.09}$ 
& $0.026^{0.010}_{0.009}$ 
& $0.13^{0.03}_{0.02}$ 
& $0.69^{0.06}_{0.07}$ 
& $0.22^{0.19}_{0.16}$ 
\\
{\small Flt+wk+LSS+SN}
& (1.00) 
& $1.05^{0.09}_{0.07}$ 
& $0.027^{0.009}_{0.009}$ 
& $0.13^{0.02}_{0.01}$ 
& $0.70^{0.05}_{0.06}$ 
& $0.20^{0.16}_{0.14}$ 
\\\hline
{\small Flt+HST}
& (1.00) 
& $1.06^{0.10}_{0.08}$ 
& $0.026^{0.010}_{0.009}$ 
& $0.15^{0.07}_{0.04}$ 
& $0.61^{0.10}_{0.21}$ 
& $0.21^{0.19}_{0.16}$ 
\\
{\small Flt+HST+LSS}
& (1.00) 
& $1.04^{0.08}_{0.07}$ 
& $0.027^{0.009}_{0.008}$ 
& $0.13^{0.02}_{0.01}$ 
& $0.68^{0.05}_{0.07}$ 
& $0.19^{0.15}_{0.13}$ 
\\
{\small Flt+HST+SN}
& (1.00) 
& $1.06^{0.11}_{0.09}$ 
& $0.027^{0.009}_{0.009}$ 
& $0.13^{0.03}_{0.02}$ 
& $0.69^{0.04}_{0.06}$ 
& $0.22^{0.19}_{0.16}$ 
\\
{\small Flt+HST+LSS+SN}
& (1.00) 
& $1.05^{0.08}_{0.07}$ 
& $0.027^{0.009}_{0.009}$ 
& $0.13^{0.02}_{0.01}$ 
& $0.70^{0.04}_{0.05}$ 
& $0.20^{0.15}_{0.14}$ 
\\\hline
\end{tabular}
\end{table}
\end{center}


The deep field measurements reveal an apparent excess in the power at
multipoles $\ell > 2000$ over standard adiabatic, inflationary models
with a significance of $3.1\sigma$. The excess is a factor of $4.5$
greater than the estimated contribution from a background of residual
sources and the confidence limit includes a $50\%$ error in the value
for the background flux density.

We have considered whether secondary anisotropies from the
Sunyaev-Zeldovich effect may explain the observed excess
\cite{Bond02}. We used four hydrodynamical simulations employing both
Smoothed Particle Hydrodynamics (SPH) and Moving Mesh Hydrodynamics
(MMH) algorithms with rms amplitudes $\sigma_8=1.0$ and $0.9$ to
calculate the expected contribution to the angular power spectrum from
the SZE. We find that both algorithms produce power consistent with
the observed excess for $\sigma_8=1$.

The \cbi\ power spectrum estimation pipeline has been tested
extensively using accurate simulations of the observations with the
exact $uv$-coverage and noise characteristics of the observed fields
\cite{Myers02}. We used the SZ maps from the hydrodynamical codes as
foregrounds in our simulations to test the Gaussian assumption
implicit in the bandpower estimation algorithm in the presence of
extended non-Gaussian foregrounds such as the SZE. We find that, for
the amplitudes considered, the pipeline recovers the total power
accurately or the 30 maps considered \cite{Bond02} including at scales
$\ell > 2000$ where the signal is dominated by the SZ foregrounds.

\section{Conclusions}
Our analysis of the \cbi\ observations has yielded parameters
consistent with the standard $\Omega_{tot}=1$, $\Lambda$CDM
model. These results, based on measurements extending to much higher
$\ell$ than previous experiments, provide a unique confirmation of the
model. The dominant feature in the data is the decline in the power
with increasing $\ell$, a necessary consequence of the paradigm which
has now been checked. In summary under weak prior assumptions the
combination of \cbi\ and \DMR\ data gives $\Omega_{tot} =
1.01_{-0.06}^{+0.09}$, and $n_s = 1.02_{-0.07}^{+0.11}$, consistent
with inflationary models; $\Omega_{\rm cdm}h^2=0.12\pm0.03$, and
$\Omega_{\Lambda} = 0.64_{-0.14}^{+0.11}$. With more restrictive
priors, flat+weak-$h$+LSS, are used, we find $\Omega_{\rm
cdm}h^2 = 0.13_{-0.01}^{+0.02}$, consistent with large scale structure
studies; $\Omega_b h^2 = 0.025_{-0.008}^{+0.010}$, consistent with Big
Bang Nucleosynthesis; $\Omega_m = 0.37\pm 0.11$, and
$\Omega_b = 0.060\pm 0.020$, indicating a low matter density universe;
$h = 0.65_{-0.12}^{+0.12}$, consistent with the recent determinations
of the Hubble Constant based on the recently revised Cepheid
period-luminosity law; and $t_0=14.0_{-1.2}^{+1.2}$ Gyr, consistent
with cosmological age estimates based on the oldest stars in globular
clusters. The combination of CMB measurements and LSS priors also
enables us to constrain the normalization $\sigma_8$. We find that for
flat+weak-$h$+LSS priors we obtain
$\sigma_8=0.89_{-0.10}^{+0.14}$. Thus it appears that the
normalization required to explain the excess with the SZE is in the
upper range of the independent result based on the primary CMB signal
and LSS data.

The 2001 observing season data is now being analyzed. Although the
data will double the overall integration time and area it is not
expected to increase the confidence of the high-$\ell$
measurements. Follow-up surveys of the deep fields in the optical
range and correlation with existing X-ray catalogs may establish
whether the measurement is indeed a serendipitous detection of the SZE
and will be part of future work. However, the observations have
highlighted the potential for SZE measurements to constrain $\sigma_8$
via the highly sensitive dependence of the angular power spectrum to
the amplitude of the fluctuations ${\cal C}^{SZ}\sim\sigma_8^7$,
although precise calibration of the theories from either numerical or
analytical methods are required to make such conclusions
feasible\cite{Bond02}. The \cbi\ is currently being upgraded with
polarization sensitive antennas for the 2002/2003 observing season.

\acknowledgements{
This work was supported by the National Science
Foundation under grants AST 94-13935, AST 98-02989, and AST
00-98734. Research in Canada is supported by NSERC and the Canadian
Institute for Advanced Research. The computational facilities at
Toronto are funded by the Canadian Fund for Innovation.  We are
grateful to CONICYT for granting permission to operate the CBI at the
Chajnantor Scientific Preserve in Chile.
}

\vfill
\end{document}